\documentclass[lettersize,journal]{IEEEtran}
\usepackage{amsmath,amsfonts}
\usepackage{algorithmic}
\usepackage{array}
\usepackage[caption=false,font=normalsize,labelfont=sf,textfont=sf]{subfig}
\usepackage{textcomp}
\usepackage{stfloats}
\usepackage{url}
\usepackage{verbatim}
\usepackage{graphicx}
\usepackage{multirow}
\usepackage{xcolor}
\usepackage{makecell}
\def\BibTeX{{\rm B\kern-.05em{\sc i\kern-.025em b}\kern-.08em
    T\kern-.1667em\lower.7ex\hbox{E}\kern-.125emX}}
\usepackage{balance}
\begin{document}
\title{Ray-Tracing Based Narrow-Beam Channel Simulation, Characterization and Performance Evaluation for 5G-R Systems}
\author{Tao Zhou,~\IEEEmembership{Senior Member,~IEEE,} Liying Geng, Kaifeng Bao, Tianyun Feng,

 Liu Liu,~\IEEEmembership{Member,~IEEE,} and Bo Ai,~\IEEEmembership{Fellow,~IEEE}
\thanks{This work was supported by the National Natural Science Foundation of China under Grants 62571024, 62221001 and 62341127. \textit{(Corresponding author: Tao Zhou.)}}
\thanks{T. Zhou, L. Geng, K. Bao, T. Feng, and L. Liu are with the School of Electronic and Information Engineering, Beijing Jiaotong University, Beijing 100044, China (e-mail: taozhou@bjtu.edu.cn;
24120045@bjtu.edu.cn; 23111017@bjtu.edu.cn; 22125022@bjtu.edu.cn; liuliu@bjtu.edu.cn).}
\thanks{B. Ai is with the State Key Laboratory of Advanced Rail Autonomous Operation, Beijing 100044, China, and also with the School of Electronic and Information Engineering, Beijing Jiaotong University, Beijing 100044, China (e-mail: boai@bjtu.edu.cn).}}
\markboth{IEEE Transactions on Vehicular Technology,~Vol.~XX, No.~YY, MONTH YEAR}%
{Shell \MakeLowercase{\textit{et al.}}: Bare Demo of IEEEtran.cls for IEEE Journals}

\maketitle

\begin{abstract}
This paper investigates narrow-beam channel characterization and performance evaluation for 5G for railway (5G-R) systems based on ray-tracing (RT) simulation. Three representative high-speed railway (HSR) scenarios including viaduct, cutting, and station are established, and RT-based dynamic narrow-beam channel simulations are conducted using a designed beam tracking scheme that ensures continuous alignment with the moving train. The channel characteristics are analyzed in terms of both large-scale and small-scale fading, as well as non-stationarity, providing statistical insights into path loss, shadow fading, fading severity, time-frequency-space dispersion, and stationarity interval. The influence of beamwidth on these channel properties is also examined. Furthermore, the performance of 5G-R systems operating in such narrow-beam channels is evaluated using the Vienna 5G simulator, with a focus on block error rate, throughput, and spectral efficiency. A hardware-in-the-loop simulation platform is developed to further assess synchronization signal reference signal received power, signal-to-interference-plus-noise ratio, and reference signal received quality. The results provide valuable guidance for the design and optimization of 5G-R systems in HSR environments.
\end{abstract}


\section{Introduction}
\IEEEPARstart{W}{ith}  the rapid maturation of fifth-generation (5G) mobile communication technology, the next-generation railway mobile communication system, 5G for railway (5G-R), has garnered significant attention in both academic and industrial domains [1], [2]. As a railway-optimized evolution of 5G technology, the 5G-R system is designed to meet high-bandwidth demands, including high-definition video transmission, real-time monitoring, and other data-intensive applications. Furthermore, it provides robust technical support for advancing railway intelligence and automation. In critical operational scenarios the ultra-low latency capabilities of 5G-R ensure reliable real-time transmission of train control commands, thereby enhancing operational safety and efficiency. As field trials and deployments progress, 5G-R is expected to play a pivotal role in elevating the quality and reliability of railway communications [3]-[5].

Massive multiple-input multiple-output (mMIMO), a key 5G technology, employs dynamic beamforming to enable highly reliable signal transmission in both sub-6 GHz and higher frequency bands. However, implementing dynamic beamforming in 5G-R systems presents a major challenge: accurate beam tracking under high-mobility conditions. While narrower beams provide higher directional gain, they also increase beam-tracking difficulty. Thus, selecting an optimal beamwidth is crucial for effective dynamic beamforming in 5G-R systems. The wireless channel is highly sensitive to variations in beam width. As the beam narrows, the number of multipath components (MPCs) decreases, altering channel characteristics. A thorough understanding of narrow-beam channel properties in railway environments is essential for applying mMIMO dynamic beamforming technology in 5G-R systems. 

Channel characterization in high-speed railway (HSR) scenarios has attracted considerable attention with several pioneering works published in the field. For HSR viaduct scenarios, authors in [6] and [7] investigated path loss (PL) and shadow fading (SF) based on the measurement results. For station scenarios, authors in [8] measured power delay profile (PDP), root-mean-square (RMS) delay spread (DS), Doppler power spectral density (DPSD) and K-factor (KF) to characterize the HSR channel. In [9], RMS angular spread (AS) and spatial characteristics were measured and analyzed for HSR channels. In [10], the non-stationary characteristic in HSR scenarios was investigated according to stationarity interval (SI). Due to the difficulty of field measurements, more researchers employ the ray-tracing (RT) method to simulate the HSR channel and analyze the channel characteristics. In [11], the PL, Doppler shifts and coherence time were studied by RT based channel simulation in different HSR scenarios such as urban, rural and tunnel. Authors in [12] studied time-varying channel characteristics in typical HSR scenarios using RT simulator. In [13], RT was used to simulate and verify KF and RMS DS of HSR channel in outdoor and tunnel environments. The above-mentioned studies have mainly concentrated on the HSR omnidirectional or wide-beam channel characteristics. 

So far, limited research has addressed narrow-beam channel characteristics, particularly in HSR scenarios. Directional antennas with different beamwidths were used in [14] for PL characterization in vehicle-to-infrastructure (V2I) highway scenarios. In [15]-[17], mMIMO channel measurements were carried out across multiple V2I settings, such as urban macrocell (UMa), rural macrocell (RMa), and urban microcell (UMi) deployments. The collected data was then used to investigate how the number of antenna elements and beam direction affect PL characteristics. Experimental characterization of narrow-beam channels was conducted in UMi, outdoor-to-indoor, and various other environments [18]-[21]. These measurements enabled quantitative analysis of beamwidth-dependent large-scale parameters, revealing an inverse correlation between antenna beamwidth and PL. Furthermore, the results demonstrated that narrower beams correspond to reduced values of both RMS DS and RMS AS. In [22], a phased-array antenna (PAA) based narrow-beam channel measurement system was utilized to conduct V2I channel measurements with different beamwidths. According to the collected data, the narrow-beam channel was characterized in terms of channel fading, time-frequency-space dispersion and non-stationarity. A geometry-based non-stationary narrow-beam channel model for high-mobility communication scenarios was proposed in [23], where the impact of half power beamwidth (HPBW) on channel statistical properties such as time-frequency-space correlation functions and DPSD was analyzed. In [24], a cluster-based dynamic narrow-beam channel model for V2I Communications was established, which considers the directional attenuation of narrow-beam and non-stationarity of clusters and provides a better characterization of the V2I narrow-beam channel. Compared to the above measurement-based and model-based approaches, RT simulation offer greater flexibility in adjusting the beamwidths of transmitting and receiving antennas and constructing the specific scenarios, thereby facilitating more comprehensive investigations into narrow-beam channel characteristics. However, the study of narrow-beam channel characteristics based on RT simulation in typical HSR scenarios is still lacking.

The existing methods for evaluating the performance of next-generation railway communication systems can be mainly categorized into three types: field testing, software simulation, and hardware-in-the-loop (HIL) simulation. Authors in [25] conducted field tests on the onboard train control system using a long-term evolution for railway (LTE-R) communication network, analyzing interference in railway scenarios and the performance of the communication system. Reference [26] introduced a design method for the future railways mobile communication system (FRMCS) in 5GRail Project and conducts field tests on the prototype. Due to the high complexity of field testing, which requires prior deployment of dedicated networks in outdoor environments, most performance evaluation studies on future railway communication systems are conducted based on software simulation or HIL simulation. Authors in [27] used the Vienna LTE-advanced system level simulator to investigate the impact of various remote unit collaboration schemes on the train average spectral efficiency in HSR scenarios. In [28], the fundamental performance of 5G-R systems considering the linear coverage scenario along the railway lines was investigated by Monte Carlo simulation, in terms of signal-to-interference-plus-noise ratio (SINR) and outage probability. In [29], the HIL simulation environment combining the actual 5G-R system and channel simulator was built, and the performance of 5G-R network bearing train control information was analyzed. Based on a HIL simulation platform with the software-defined radio (SDR) and channel emulator, the physical layer transmission performance of the 5G-R communication system was evaluated by bit error rate (BER) and error vector magnitude (EVM) [30]. In current 5G-R performance simulations, standard channel models such as 3GPP TR 38.901, WINNER II, and IMT-2020 are predominantly used. However, these models cannot reflect narrow-beam channel characteristics in specific HSR scenarios, which will influence the accuracy of performance evaluation for 5G-R systems in case of narrow-beam. 

To fill the aforementioned research gaps, this paper aims to characterize the narrow-beam channel in typical HSR scenarios based on RT simulation and evaluate the performance of 5G-R systems in narrow-beam channels. The major contributions and novelties of this paper are as follows.
\begin{enumerate}
\item{Three typical HSR scenarios including viaduct, cutting and station are constructed and RT-based narrow-beam channel simulation in these scenarios is performed. A beam tracking scheme is designed to achieve the dynamic simulation, which enables the beam direction to be constantly aligned with the moving train.}
\item{The narrow-beam channel in the HSR scenarios is characterized in terms of large-scale and small-scale fading and non-stationarity. The statistical results related to PL, SF, fading severity, time-frequency-space dispersion and SI are presented, and the impact of beamwidth on the channel characteristics is analyzed.}
\item{The performance of 5G-R systems in HSR narrow-beam channels is evaluated based on the Vienna 5G simulator, focusing on block error rate (BLER), throughput and spectral efficiency. A 5G-R HIL simulation platform is established and further used to evaluate the performance in terms of synchronization signal (SS) reference signal received power (RSRP), SINR and reference signal received quality (RSRQ).}
\end{enumerate}

The remainder of this paper is organized as follows. Section II describes the RT-based narrow-beam channel simulation for the three HSR typical scenarios. In Section III, the major characteristics of narrow-beam channel for the HSR scenarios are discussed. Then, the performance evaluation for 5G-R systems in HSR narrow-beam channels is introduced in Section IV. Finally, conclusions are drawn in Section V.   

\section{RT-Based Narrow-Beam Channel Simulation for HSR Scenarios}

\subsection{Scenario Construction}

In order to achieve HSR channel simulation, we consider three typical HSR scenarios, including viaduct, cutting and station, which are constructed using Google SketchUp. Fig. 1 shows the established HSR scenarios, and the detail description is introduced as follows.

\begin{figure*}[!t]
\centering
\vspace{-0.45cm}
\captionsetup[subfigure]{font=footnotesize,labelfont=rm,textfont=rm}
\subfloat[]{\includegraphics[scale=0.35]{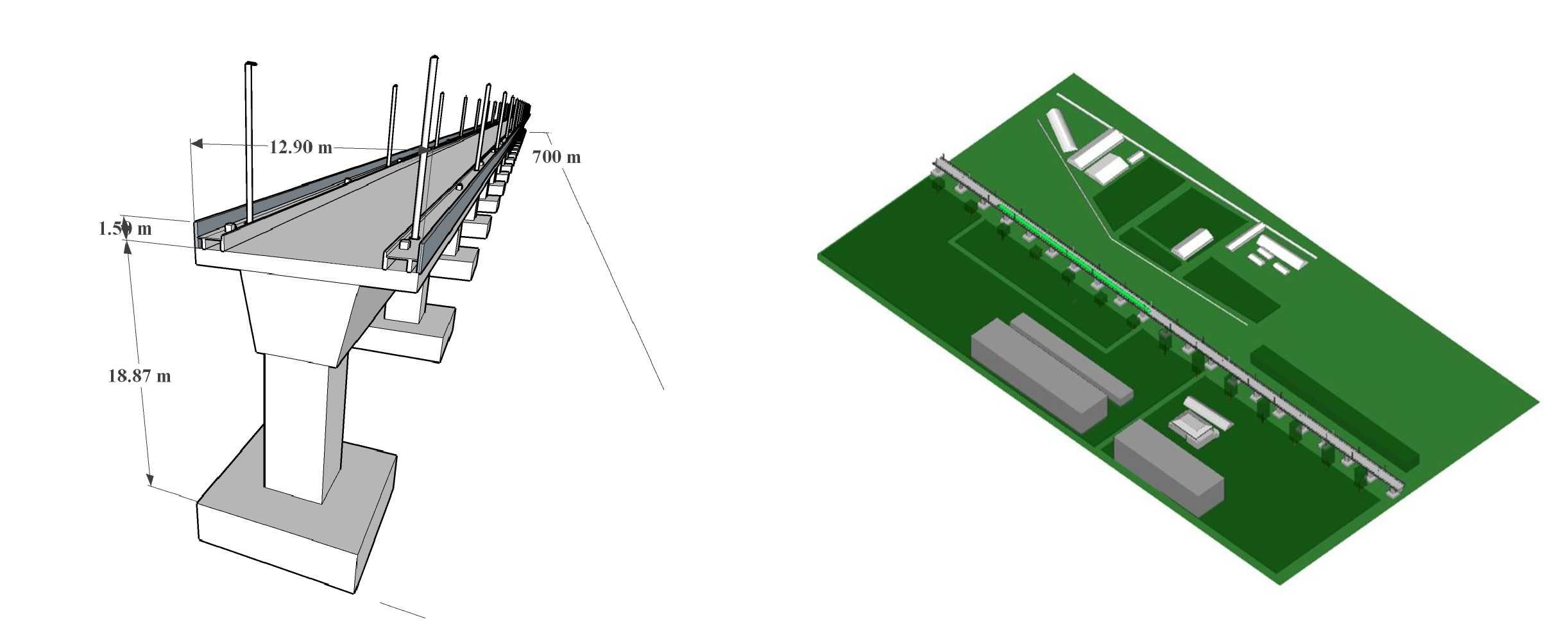}%
    \label{fig_second_case}}\\
\vspace{-0.45cm}
\subfloat[]{\includegraphics[scale=0.7]{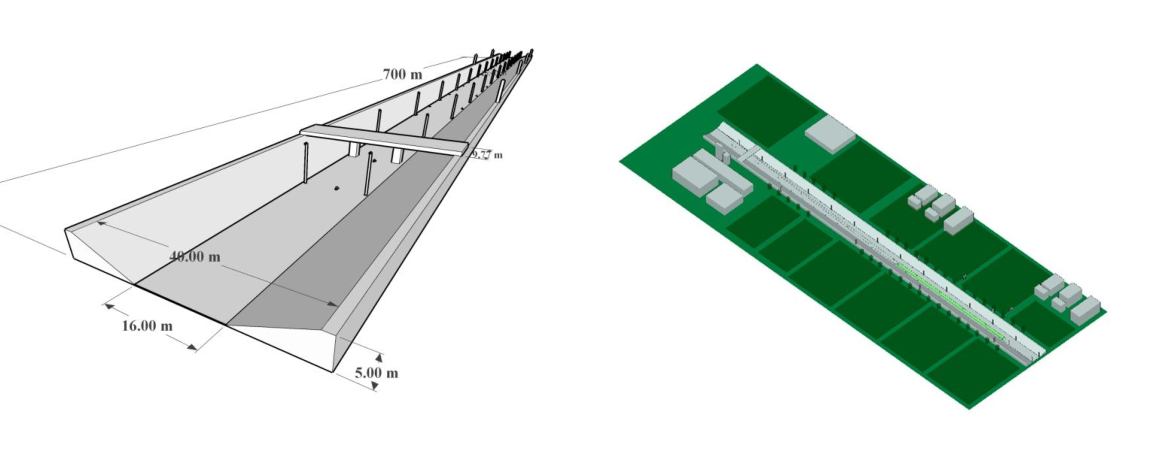}%
	\label{fig_second_case1}}\\
\vspace{-0.45cm}
\subfloat[]{\includegraphics[scale=0.4]{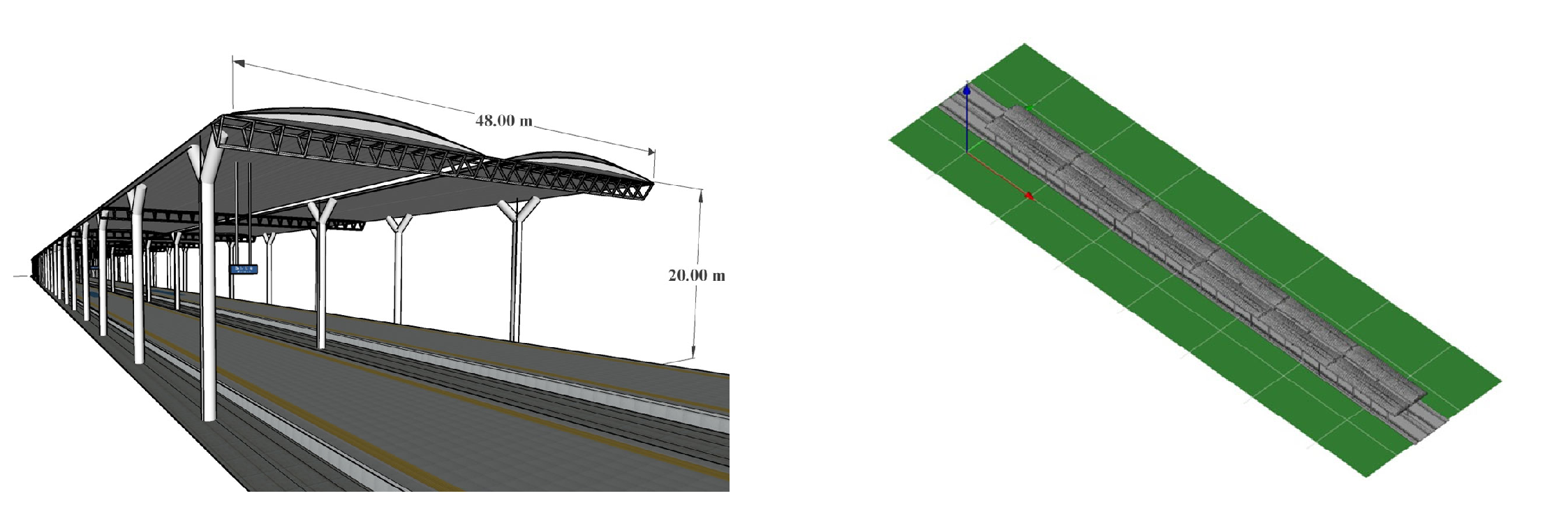}%
	\label{fig_second_case3}}
\caption{Constructed HSR scenarios. (a) Viaduct. (b) Cutting. (c) Station.}
\label{fig_sim}
\vspace{-0.35cm}
\end{figure*}
\textit{1) Viaduct:} Viaduct is a long bridge type structure and usually with series of arches, whose role is to carry the railway in a valley or other uneven ground. The viaduct is usually 10-30 m high. The transmitter (TX) antenna is usually 20-30 m higher than the track surface. The receiver (RX) antenna is installed on top of trains. In viaduct scenarios, few loose reflectors can exceed the height of the bridge, so the line-of-sight (LOS) propagation mode of radio waves dominates. The type of the constructed viaduct in this paper refers to Viaduct 1a mentioned in [31]. As shown in Fig. 1(a), there are few tall trees and low bushes higher than the viaduct surface. Farms, schools and billboards are distributed on both sides of the viaduct. The bridge height is 19 m, the top width is 13 m, and the guardrail is 1.5 m. The size of the constructed viaduct scenario is 700 m × 400 m.

\textit{2) Cutting:} Cutting is to lay railway on uneven ground, so that the high-speed train can pass through large obstacles such as mountains. To prevent the main building from sinking, the side walls of the cutting are covered with vegetation and reinforced concrete. The constructed cutting scenario in this paper refers to the size of U-shaped groove in the field measurement in literature [32], as shown in Fig. 1(b). The depth, bottom width and top width of the cutting are 5 m, 16 m and 40 m, respectively, and the narrow bridge width at top of the cutting is 9.77 m. Poles are evenly arranged along railway trackside with 9.3 m height. Cable boxes and track boxes are distributed in trackside space. Trees and coppice are distributed on both sides of the cutting, among which the highest trees are 10 m, the highest coppice is 3 m, and the highest buildings are 20 m. The size of the constructed cutting scenario is 700 m × 400 m.

\textit{3) Station:} Station, as a hub of railway transportation, serves the function of passengers boarding and alighting from trains, which is usually composed of platforms and ceilings. The speed of high-speed train in medium and large stations is often lower than 80 km/h, while through small stations often stays the same. Therefore, when studying channel characteristics over 200 km/h, the construction of the station scenario should not be too complicated. As shown in Fig. 1(c), the type of constructed station scenario in this paper refers to Station-4A mentioned in [31], and the building construction and surrounding environment refer to Yinchuan East Station in China. There exist LOS and none-line-of-sight (NLOS) propagation cases in the constructed station scenario, where the scatterers mainly include platforms, ceilings, columns, station signs, etc. The platform is 600 m long, 48 m wide, and the ceiling is 20 m high. The size of constructed station scenario is 700 m × 300 m.

\vspace{-0.35cm}
\subsection{Simulation Configuration}

According to the constructed HSR scenarios, we perform narrow-beam channel simulation using a commercial RT software, Wireless Insite, developed by REMCOM. In the RT simulation, the center frequency and bandwidth are chosen as 2.1 GHz and 10 MHz, respectively, which are consistent with the candidate frequency band of 5G-R systems in China. At TX side, in addition to an omnidirectional antenna, three directional antennas with different beam types are considered, including Type A (60° horizontal HPBW and 10° vertical HPBW), Type B (30° horizontal HPBW and 10° vertical HPBW), and Type C (12° horizontal HPBW and 10° vertical HPBW). The Type A antenna belongs to wide-beam antenna while the Type B and Type C antennas can be regarded as narrow-beam antenna. Based on these different TX antennas, we can investigate the impact of beamwidth on channel characteristics. The distance between the TX antenna and the railway track is set to 100 m, and the height of the TX antenna is set to 22 m higher than the railway track. At RX side, an omnidirectional antenna is employed and is set to 3.1 m higher than the railway track, and the RX is placed at 1 m intervals.

The electromagnetic parameters of different materials determine the transmission power of electromagnetic waves with reflection, diffraction and scattering. There are six kinds of materials used in our simulation, including concrete, metal, marble, soil, trunk and leaf. Table I lists the relative permittivity and conductivity for each material. The electromagnetic parameters selected for different materials refer to ITU-R P.2040-2 [33]. In addition, the maximum numbers of reflection and diffraction are set to 6 and 1, respectively, and the scattering model is chosen as Lambert model.

\begin{table}[t]
\renewcommand\arraystretch{1.2}
\small \caption{Configuration of electromagnetic parameters of different materials.} \label{table1} \centering
\begin{tabular}{c | c | c}
\hline\hline
\textbf{Material}	&\textbf{Relative permittivity}   &\textbf{Conductivity(S/m)}\\
\hline
Concrete & 5.31 & 0.06622\\
\hline
Metal & 1 & $10^{7}$\\
\hline
Marble & 7.04 & 0.93\\
\hline
Soil & 13.74 & 0.14\\
\hline
Trunk & 1.99 & 0.01201\\
\hline
Leaf  & 20 & 0.39\\
\hline
\end{tabular}
\vspace{-0.45cm}
\end{table}

\vspace{-0.3cm}
\subsection{Dynamic Simulation}

In the case of narrow-beam antenna, we should perform a dynamic simulation with the consideration of beam tracking. Unlike the wide-beam antenna, the beam direction of the narrow-beam antenna needs to be constantly aligned with the moving train, in order to avoid coverage holes and poor coverage performance. In the simulation, the beam tracking is achieved by updating the beam direction at certain intervals, as shown in Fig. 2. It is noted that the updating interval should be less than the distance of coverage area $(s_{0},s_{1},\ldots,s_{n})$ of narrow-beam antenna, which is determined by the beam direction and beamwidth. 
\begin{figure}[!t]
\centering
\vspace{0.15cm}
\includegraphics[width=3.3in]{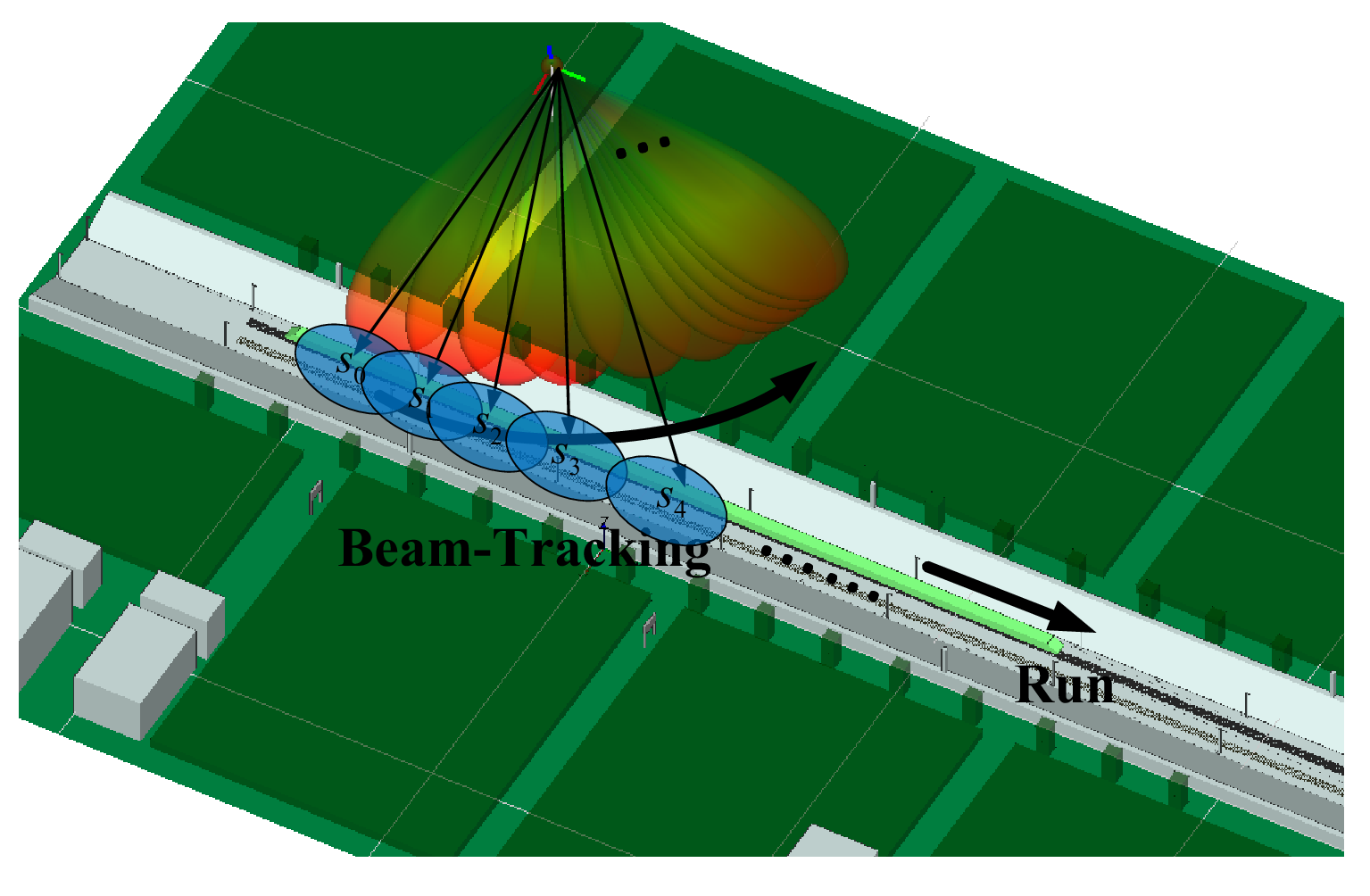}
\caption{Beam tracking scheme.}
\vspace{-0.15cm}
\label{fig2}
\end{figure}

\begin{figure}[!t]
\vspace{-0.25cm}
\centering
\includegraphics[width=3in]{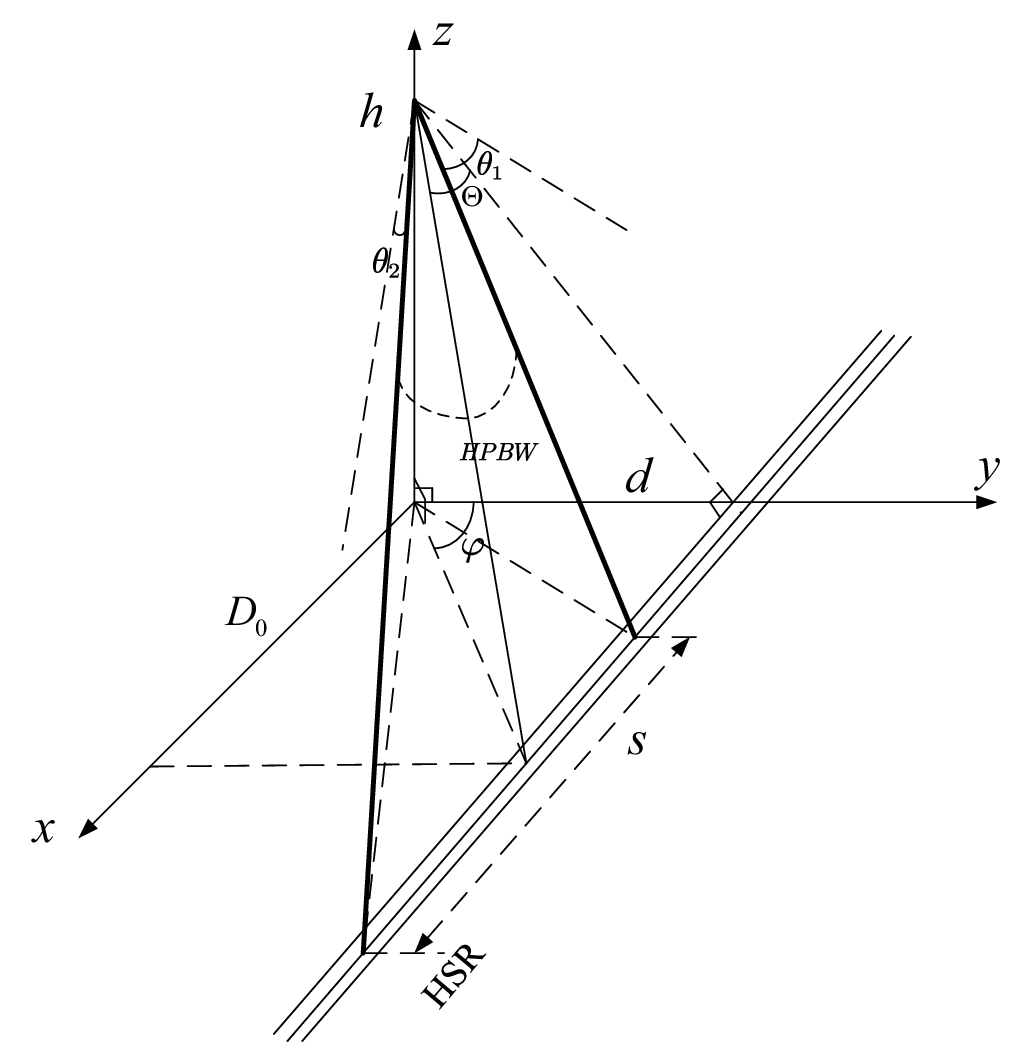}
\caption{A geometrical model for calculating the coverage distance.}
\label{fig3}
\vspace{-0.6cm}
\end{figure}

To calculate the coverage distance, a geometrical model with narrow-beam antenna in the HSR scenario is considered, as shown in Fig. 3. The beam direction $\Theta$ and coverage distance $s$ can be expressed as
\begin{equation}\label{eq:1}
\Theta = \cos^{-1} \frac{\sqrt{h^2 + d^2}}{\sqrt{h^2 + \left( \frac{d}{\cos\varphi} \right)^2}} = \tan^{-1} \frac{D_0}{\sqrt{h^2 + d^2}},
\end{equation}
\begin{equation}\label{eq:2}
\begin{split}
s &= \sqrt{h^2 + d^2} \left( \tan(\Theta + HPBW / 2)\right.\\
&\quad\left.-\ tan(\Theta - HPBW / 2) \right),
\end{split}
\end{equation}
where $h$ is the antenna height, $d$ denotes the distance between the antenna and the railway track, $\varphi$ represents the beam direction in horizontal dimension, $D_{0}$ stands for the distance between the train and the antenna on x-axis, and $HPBW$ is the value of HPBW. Generally, the $d$ is much larger than the $h$ in HSR scenarios, which meets the following requirements
\begin{equation}\label{eq:3}
\vspace{0.3cm}
\left\{
\begin{aligned}
\sqrt{h^2 + d^2} &\approx d \\
\Theta &\approx \varphi.
\end{aligned}
\right.
\end{equation}

Therefore, Eq. (2) can be simplified as
\begin{equation}\label{eq:4}
s = d \left( \tan(\varphi + HPBW / 2) - \tan(\varphi - HPBW / 2) \right).
\end{equation}

From Eq. (4), the coverage distance of a specific horizontal beamwidth under different beam directions can be calculated. When the horizontal HPBW is 12° or 30°, the beam direction should be updated every 20 m or 50 m, which can meet the requirement of beam tracking in our simulation. Owing to the complexity of beamforming algorithms, next-generation antenna arrays for railway base stations will still not incorporate vertical beamforming capability. Thus, only horizontal beam tracking is required in this study.

\section{Narrow-Beam Channel Characterization for HSR Scenarios}

\subsection{Large-Scale Fading Characteristics}

Based on the results of narrow-beam channel simulation for HSR scenarios, we firstly analyze large-scale fading characteristics of the HSR narrow-beam channel in terms of PL and SF.

\begin{figure*}[!t]
\centering
\vspace{-0.45cm}
\includegraphics[width=0.7\linewidth]{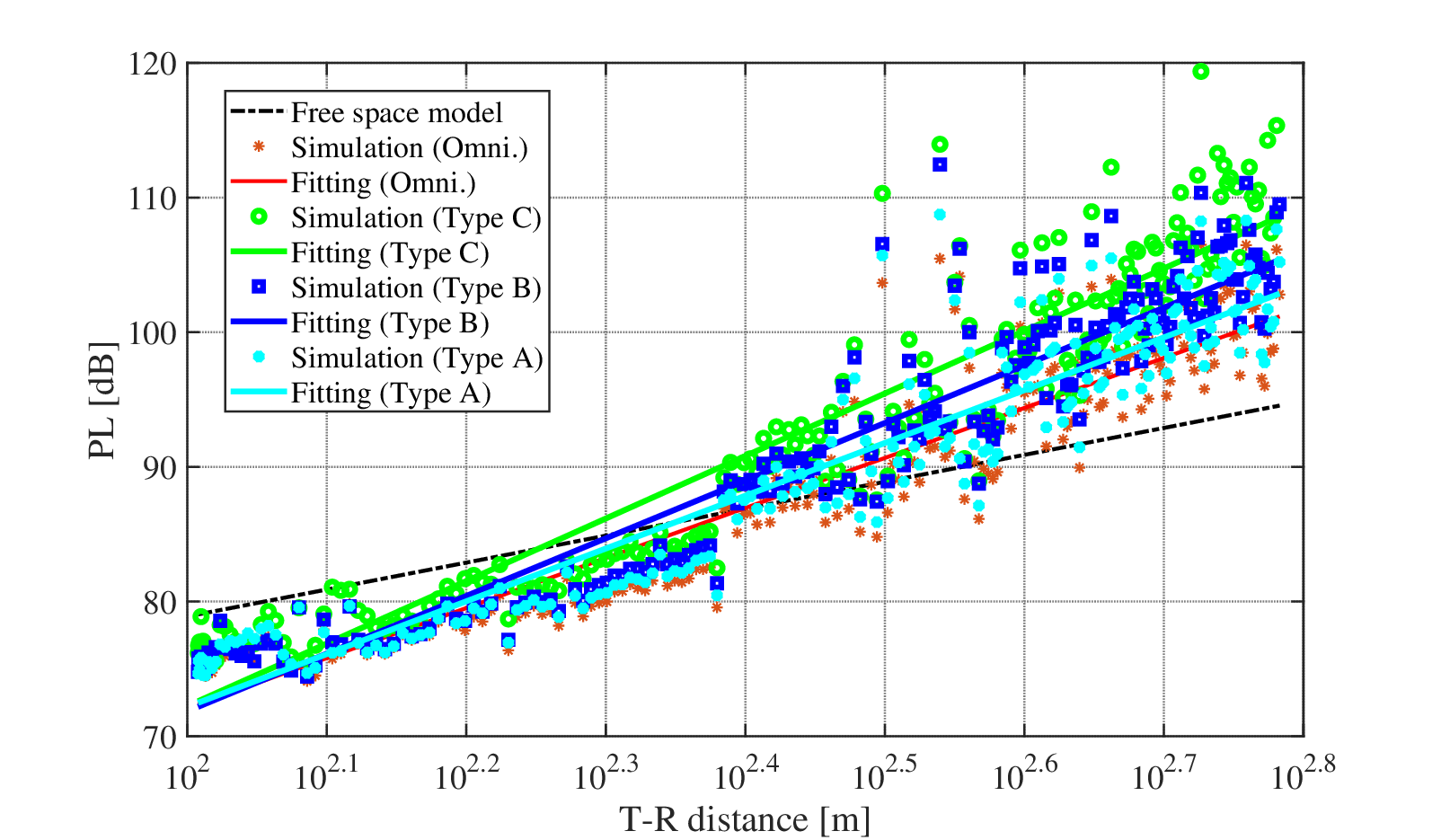}
\caption{PL results in the viaduct scenario.}
\label{fig4}
\vspace{-0.35cm}
\end{figure*}

\textit{1) PL:} We obtain the PL results in the three HSR scenarios with four kinds of antenna configuration and apply a classical Logarithmic distance model to fit the results using the least square (LS) method. As an example, Fig. 4 illustrates the PL values and fitted PL curves in the viaduct scenario. Note that the result of free space model is also provided for comparison. It can be found that the fitted PL curves against the distance between TX and RX (T-R distance) are lower than the curve of free space model in the initial distance. This is because the simulated MPCs have small phase difference within a relatively close distance, which leads to the increase of received power. When the distance is greater than a certain distance, the fitted PL curves are gradually higher than the curve of free space model. It can be also seen that from the narrow-beam antenna (Type C) to the omnidirectional antenna, the PL decreases gradually at the same distance. This is due to the fact that as the beamwidth increases, more MPCs can be identified and higher received power can be achieved. It should be mentioned that the above findings are similar in the three HSR scenarios.

Table II compares results of PL exponent $n_{PL}$ in the three HSR scenarios with four kinds of antenna configuration. It is observed that the $n_{PL}$ for the Type B and Type C is higher than that for the Type A and omnidirectional antenna, which means the narrow-beam antenna causes the larger PL. Nevertheless, this shortcoming can be compensated by the higher antenna gain of the narrow-beam antenna. In addition, it is found that for the four kinds of antenna configuration, the station scenario has much lower $n_{PL}$ than the viaduct and cutting scenarios. The reason is that a large number of MPCs are produced by the rich scatterers in the station scenario, which increases the received power.

\begin{table*}[t]
  \renewcommand\arraystretch{1.8} 
  \centering
  \vspace{-0.15cm}
  \small\caption{Parameters of channel characteristics in simulated HSR scenarios}
  \begin{tabular}{>{\centering\arraybackslash}m{1.3cm}|>{\centering\arraybackslash}m{0.8cm}| *{4}{>{\centering\arraybackslash}m{0.8cm}}| *{4}{>{\centering\arraybackslash}m{0.8cm}}| *{4}{>{\centering\arraybackslash}m{0.8cm}}}
    \hline\hline
    \multicolumn{2}{c|}{\textbf{Scenario}} & \multicolumn{4}{c|}{\textbf{Viaduct}} & \multicolumn{4}{c|}{\textbf{Cutting}} & \multicolumn{4}{c}{\textbf{Station}} \\
    \hline
    \multicolumn{2}{c|}{Beam Type} & Omni. & A & B & C & Omni. & A & B & C & Omni. & A & B & C \\
    \hline
    \multirow{2}{*}{\makecell{PL {[dB]}\\~}} & $n_{\text{PL}}$ & 3.70 & 3.90 & 4.60 & 4.70 & 3.90 & 4.10 & 4.47 & 4.84 & 3.07 & 3.46 & 3.48 & 3.68 \\
    \hline
    \multirow{2}{*}{\makecell{SF {[dB]}\\~}} & $\sigma_{\text{SF}}$ & 3.30 & 3.31 & 3.49 & 4.18 & 4.57 & 4.76 & 4.90 & 5.13 & 4.31 & 4.89 & 4.74 & 4.78 \\
    \hline
    \multirow{2}{*}{\makecell{KF {[dB]}}} & $\mu_K$ & 4.96 & 5.23 & 5.40 & 5.43 & 2.81 & 2.65 & 3.14 & 3.65 & 2.06 & 3.06 & 4.06 & 4.77 \\
    & $\sigma_K$ & 4.56 & 5.11 & 4.30 & 4.33 & 4.50 & 4.24 & 3.94 & 3.99 & 5.07 & 3.49 & 2.04 & 2.16 \\
    \hline
    \multirow{2}{*}{\makecell{DS {[ns]}}} & $\mu_{DS}$ & 40.60 & 35.58 & 25.62 & 14.11 & 127.09 & 97.84 & 73.96 & 15.96 & 93.51 & 59.37 & 53.43 & 41.87 \\
    & $\sigma_{DS}$ & 33.62 & 29.88 & 19.13 & 12.44 & 133.31 & 144.41 & 103.20 & 21.02 & 34.56 & 30.89 & 12.64 & 11.61 \\
    \hline
    \multirow{2}{*}{\makecell{DPS {[Hz]}}} & $\mu_{DPS}$ & 69.11 & 67.72 & 52.63 & 44.09 & 113.84 & 102.57 & 99.55 & 47.74 & 170.67 & 159.26 & 152.86 & 148.40 \\
    & $\sigma_{DPS}$ & 47.99 & 43.35 & 40.94 & 41.51 & 35.33 & 33.22 & 28.37 & 27.81 & 122.08 & 126.84 & 121.24 & 115.84 \\
    \hline
    \multirow{2}{*}{\makecell{AAS {[deg]}}} & $\mu_{AAS}$ & 15.60 & 13.45 & 12.44 & 12.46 & 23.11 & 18.36 & 14.78 & 9.12 & 29.01 & 23.11 & 22.01 & 21.28 \\
    & $\sigma_{AAS}$ & 10.40 & 8.73 & 5.97 & 5.89 & 20.83 & 14.87 & 12.39 & 8.81 & 20.83 & 14.83 & 11.98 & 10.34 \\
    \hline
    \multirow{2}{*}{\makecell{DAS {[deg]}}} & $\mu_{DAS}$ & 2.31 & 2.34 & 1.64 & 1.33 & 2.27 & 1.95 & 1.58 & 1.32 & 15.90 & 7.23 & 3.61 & 1.92 \\
    & $\sigma_{DAS}$ & 1.65 & 0.76 & 0.68 & 0.45 & 2.61 & 1.13 & 1.45 & 0.75 & 6.70 & 2.95 & 1.57 & 1.01 \\
    \hline
  \end{tabular}
  \label{table2}
  \vspace{-0.45cm}
\end{table*}

\textit{2) SF:} Based on the PL results, we further obtain the standard deviation of SF ($\sigma_{SF}$). Table II compares the results of $\sigma_{SF}$ in the three HSR scenarios with four kinds of antenna configuration. It can be seen that in each scenario the $\sigma_{SF}$ tends to increase with the decrease of beamwidth, but the change is small. This is because when there is a building between the TX and the RX, the multipath signal transmitted by a narrow-beam antenna is mostly blocked, and only a small part of the signal arrives at the RX through diffraction and transmission. When the RX moves, the relative angle between it and the scatterer changes, so the transmission and diffraction coefficient change, which thus intensifies the variation of superposition field strength at the RX side. In addition, by comparing the $\sigma_{SF}$ of each antenna type across the three HSR scenarios respectively, it can be found that the $\sigma_{SF}$ is the largest in the cutting scenario, followed by the station scenario, and the viaduct scenario is significantly lower than the former. This is due to the semi-closed structure of the cutting and station scenarios, which is stronger to shield the signal. The viaduct scenario is LOS dominant, and there are few trees that can exceed the height of the viaduct, so the $\sigma_{SF}$ is relatively small.

\vspace{-0.45cm}
\subsection{Small-Scale Fading Characteristics}

The small-scale fading characteristics of the HSR narrow-beam channel can be divided into fading severity and time-frequency-space dispersion, which will be analyzed as follows.

\textit{1) Fading Severity:} The small-scale fading severity is usually quantified by the KF, which can be directly calculated according to the power ratio of the dominant MPC and the other MPCs. As an example, Fig. 5 illustrates the KF result in the cutting scenario with the narrow-beam antenna (Type C). It can be seen that the KF value shows little variation within the 0-150 m range but gradually increases beyond 150 m. This is because, at closer distances to the TX, a significant portion of the signal is obstructed by trees near the cutting. As the angle changes, the obstructive effect of the trees on the signal weakens, leading to an increase of the KF. Additionally, due to the U-shaped groove structure of the cutting, the top of the cutting obstructs the signal near the TX position perpendicular to the cutting. The LOS path cannot bypass the top of the cutting to reach the RX directly, resulting in a higher proportion of scattered, reflected, and diffracted components from the inner walls of the trench in the received signal at closer distances, which lowers the KF. In the far-end coverage area of the TX (200-400 m), the influence of the groove on the signal weakens, and the KF fluctuates between 5-15 dB. From 420 m to 500 m, the KF remains relatively stable because there is no vegetation obstruction in this section of the scenario. Furthermore, at 530 m, the KF drops significantly, and shows a decrease followed by an increase between 530 m and 600 m. This is due to the narrow bridge blocking the LOS propagation of the signal, leading to a lower KF intensity.

\begin{figure}[!t]
\centering
\vspace{-0.31cm}
\includegraphics[width=3.5in]{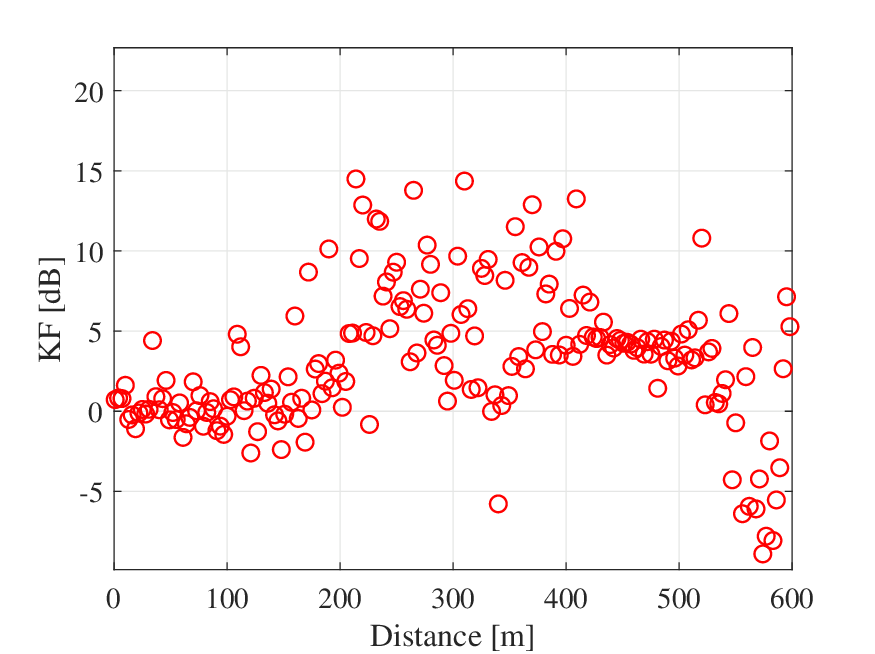}
\caption{KF result in the cutting scenario with the narrow-beam antenna (Type C).}
\label{fig5}
\vspace{-0.7cm}
\end{figure}

Table II lists the KF results in terms of mean value $\mu_{K}$ and standard deviation $\sigma_K$ in the three HSR scenarios with four kinds of antenna configuration. It can be seen that the $\mu_K$ in the viaduct scenario is the largest. This is because the height of viaduct is significantly higher than that of most trees and buildings and there is less occlusion of the RX by scatterers and reflectors, resulting in the weaker small-scale fading. It can be also found that as the beamwidth decreases, $\mu_K$ gradually increases while $\sigma_K$ shows a decreasing trend, and this decreasing trend is more pronounced in the station scenario. This is due to the reduced beamwidth, which significantly decreases the number of MPCs, leading to a higher energy dominance of the main MPC.

\begin{figure*}[!t]
\centering
\vspace{-0.6cm}
\captionsetup[subfigure]{font=footnotesize,labelfont=rm,textfont=rm}
\subfloat[]{\includegraphics[width=0.45\linewidth]{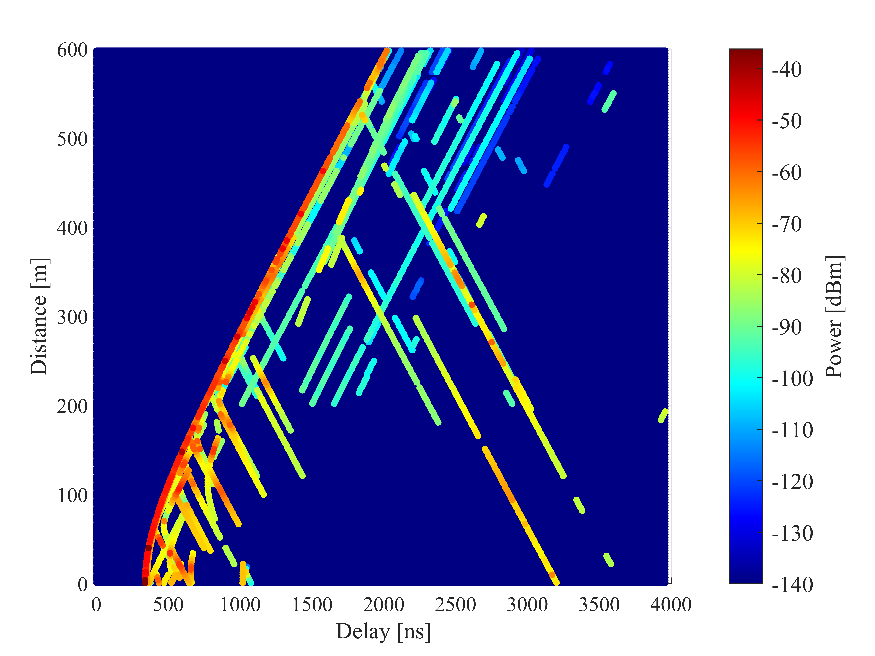}%
\label{fig_first_case}}
\hfil
\subfloat[]{\includegraphics[width=0.45\linewidth]{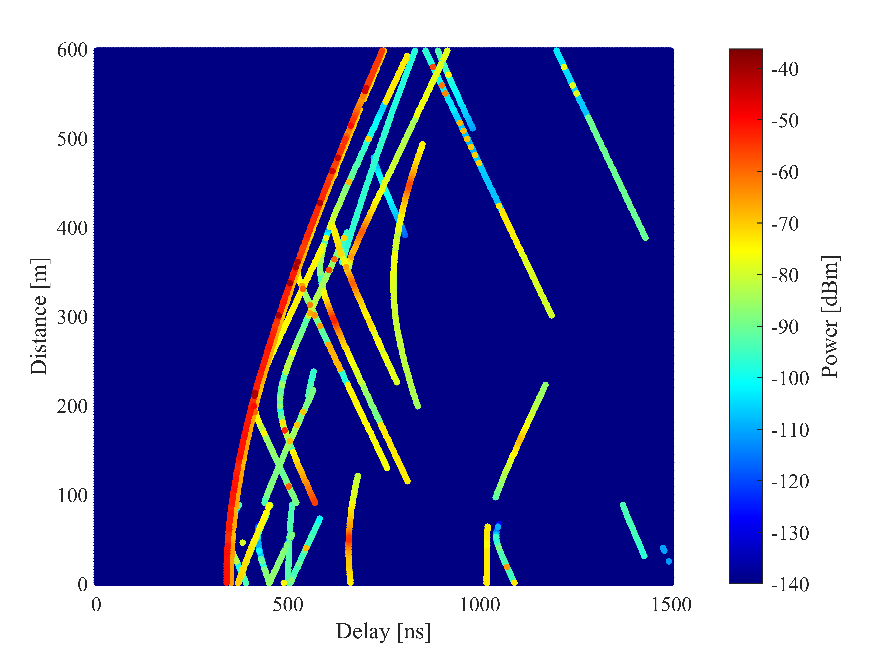}%
\label{fig_second_case}}
\caption{PDP results in the cutting scenario. a) Omnidirectional antenna. b) Narrow-beam antenna (Type C).}
\label{fig_sim}
\vspace{-0.45cm}
\end{figure*}

\textit{2) Time Dispersion:} The time dispersion can be quantified by the RMS DS, which is calculated according to the PDP. Fig. 6 shows PDP results in the cutting scenario with the omnidirectional antenna and narrow-beam antenna (Type C). It is observed that the PDP contains several MPCs that are parallel to the main path and exhibit very small relative delay variations. These MPCs primarily originate from single or multiple reflections on the cutting sidewalls. The single-reflection MPCs exhibit higher power and are closely clustered around the main path, whereas the multiple-reflection MPCs have lower power and are distributed farther from the main path. Additionally, such MPCs predominantly appear beyond 200 m. This is because, as the train moves away from the TX, the angle between the beam direction and the normal vector of the cutting increases, causing more signals to reach the RX via reflections within the cutting. It can be also seen that the PDP contains another type of MPCs that appears as “spikes” with a certain angle relative to the main path. These MPCs partly originate from distant buildings or scatterers far from the railway track, and partly from utility poles alongside the track. Such MPCs exhibit relatively large delays and low power levels. A comparison between the PDP results for the omnidirectional antenna and narrow-beam antenna (Type C) reveals that the first type of MPCs still exists, while the second type nearly disappears. This is because the narrow-beam antenna exhibits stronger directivity, making it difficult for MPCs to reach the RX through multiple distant reflections. Thus, the narrow-beam antenna can mitigate the impact of the cutting structure on wireless signal transmission. Additionally, while the number of the second type of MPCs shows little difference between the narrow-beam and omnidirectional channels, it can be observed that for the omnidirectional channel, MPCs from certain scatterers can persist for up to approximately 1500 ns, whereas for the narrow-beam channel the maximum MPC duration is only about 500 ns. This indicates that the MPC lifetime in the narrow-beam channel is significantly shorter compared to the omnidirectional channel.

Table II lists the mean $\mu_{DS}$ and standard deviation $\sigma_{DS}$ of RMS DS results. Among the three HSR scenarios, the omnidirectional antenna exhibits the largest $\mu_{DS}$, followed by Type A and Type B, with Type C shows the smallest $\mu_{DS}$. This indicates that as the beamwidth increases, the channel experiences more severe time dispersion, resulting in larger RMS DS. In the cutting scenario, the $\mu_{DS}$ of the omnidirectional channel is significantly higher than that of narrow-beam channels. This is attributed to the fact that the omnidirectional channel incorporates a large number of MPCs reflected by the U-shaped groove before reaching the RX, thereby increasing the mean excess delay. By contrast, the narrow-beam channel is minimally impacted by the U-shaped groove, leading to time dispersion that is significantly weaker than that of the omnidirectional channel.

\begin{figure*}[!t]
\vspace{-0.6cm}
\centering
\captionsetup[subfigure]{font=footnotesize,labelfont=rm,textfont=rm}
\subfloat[]{\includegraphics[width=0.45\linewidth]{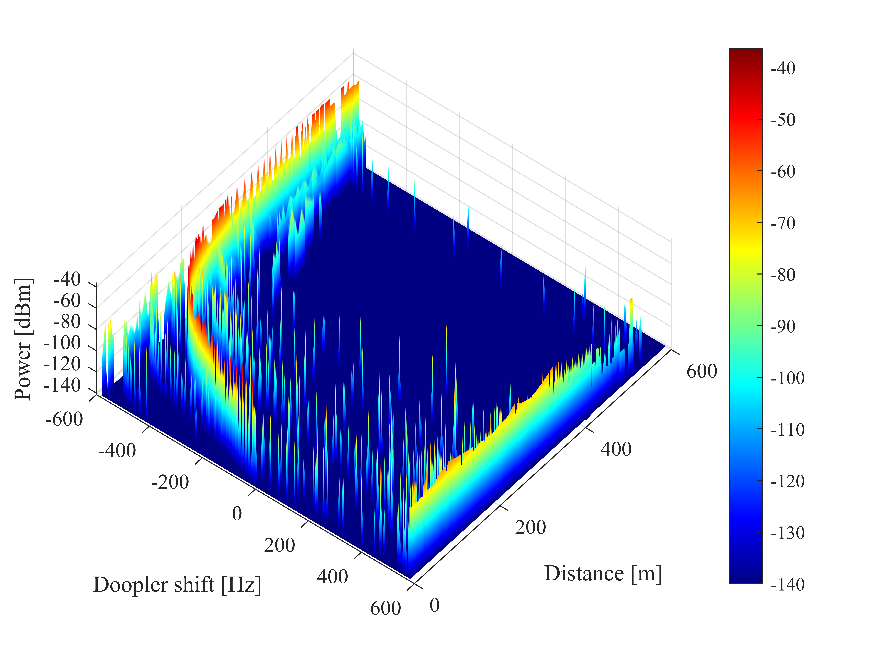}%
\label{fig_first_case}}
\hfil
\subfloat[]{\includegraphics[width=0.45\linewidth]{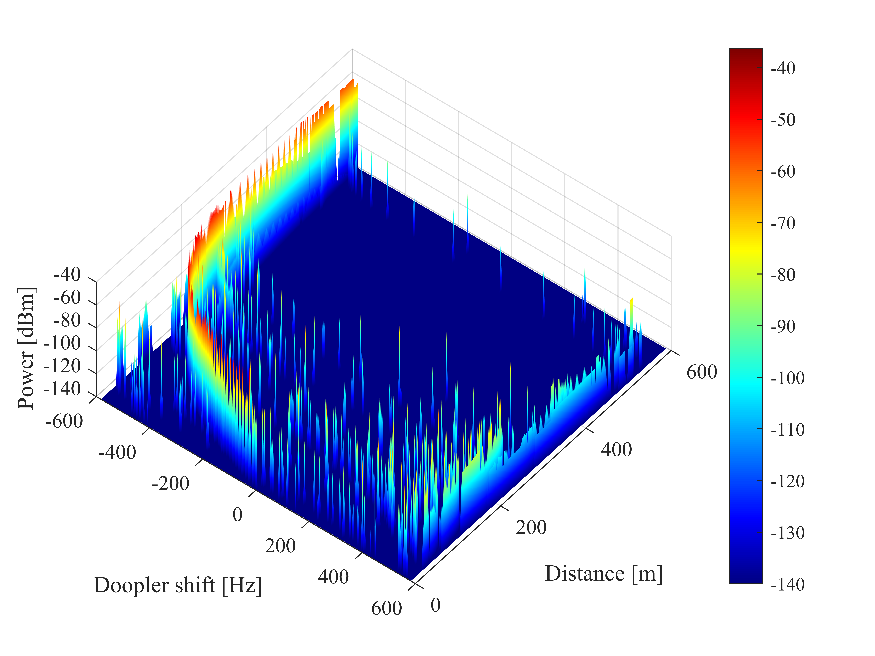}%
\label{fig_second_case}}
\caption{DPSD results in the cutting scenario. a) Omnidirectional antenna. b) Narrow-beam antenna (Type C).}
\label{fig_sim}
\vspace{-0.45cm}
\end{figure*}

\textit{3) Frequency Dispersion:} In order to quantitatively describe the frequency dispersion, the RMS Doppler spread (DPS) can be used, which is computed from the DPSD. Fig. 7 illustrates DPSD results in the cutting scenario with the omnidirectional antenna and directional antenna (Type C). It is observed that when the train travels to 600 m, the Doppler shift reaches the maximum of 570 Hz, which is in agreement with the theoretical value when the velocity of the is set to 300 km/h. A comparison between the DPSD results for the omnidirectional antenna and narrow-beam antenna (Type C) reveals the following findings. For the narrow-beam antenna, Doppler shifts are primarily concentrated in the range of $-$500 to 0 Hz and cluster around the Doppler shift curve of the main path. In contrast, the omnidirectional antenna exhibits more dispersed frequency offsets. Notably, the DPSD of the omnidirectional channel consistently displays a spectral line near 500 Hz (corresponding to the maximum Doppler shift), which is virtually absent in the narrow-beam channel. This occurs because the omnidirectional antenna’s broader radiation pattern enables continuous reception of MPCs with angles nearly parallel to the direction of motion, resulting in a persistent maximum-Doppler-shift spectral line in the DPSD.

Table II lists the mean $\mu_{DPS}$ and standard deviation $\sigma_{DPS}$ of RMS DPS results. It can be seen that as the beamwidth decreases, both the $\mu_{DPS}$ and $\sigma_{DPS}$ exhibit a decreasing trend. Among all scenarios, the cutting scenario displays the most significant impact of beamwidth variation on frequency dispersion. Furthermore, the viaduct scenario exhibits the smallest $\mu_{DPS}$, while the station scenario shows the largest $\mu_{DPS}$ under the same beamwidth.

\textit{4) Space Dispersion:} The RMS AS serves as a metric to quantify the space dispersion of the channel, which can be derived from the power angular spectrum (PAS). Fig. 8 illustrates PAS results of angle of departure (AoD) and angle of arrival (AoA) in the cutting scenario with the directional antenna (Type C). It is found that there exists a distinct main path, which reflects the train’s movement trajectory. When the train approaches the TX, the AoD reaches its maximum value of 90°; conversely, when the train moves away, the AoD decreases to its minimum of 10°. As for the AoA, it reaches its minimum of $-$90° when the train approaches the TX, while it peaks at $-$10° when the train moves away from the TX. When the train moves infinitely far from the TX, this angle converges to 0°. The simulation results align well with the geometric relationships.

Table II lists the mean $\mu_{AAS}$ and standard deviation $\sigma_{AAS}$ of RMS AS of AOA (AAS) results and the mean $\mu_{DAS}$ and standard deviation $\sigma_{DAS}$ of RMS AS of AOD (DAS) results. It can be seen that the scheme with a wider beamwidth exhibits larger AS. The RMS AAS for the narrow-beam antenna (Type C) in the station scenario is 21.28°, significantly higher than the values of 9.12° and 12.46° observed in the cutting and viaduct scenarios, respectively. This indicates that the station scenario experiences stronger spatial dispersion.

\begin{figure*}[!t]
\centering
\vspace{-0.6cm}
\captionsetup[subfigure]{font=footnotesize,labelfont=rm,textfont=rm}
\subfloat[]{\includegraphics[width=0.48\linewidth]{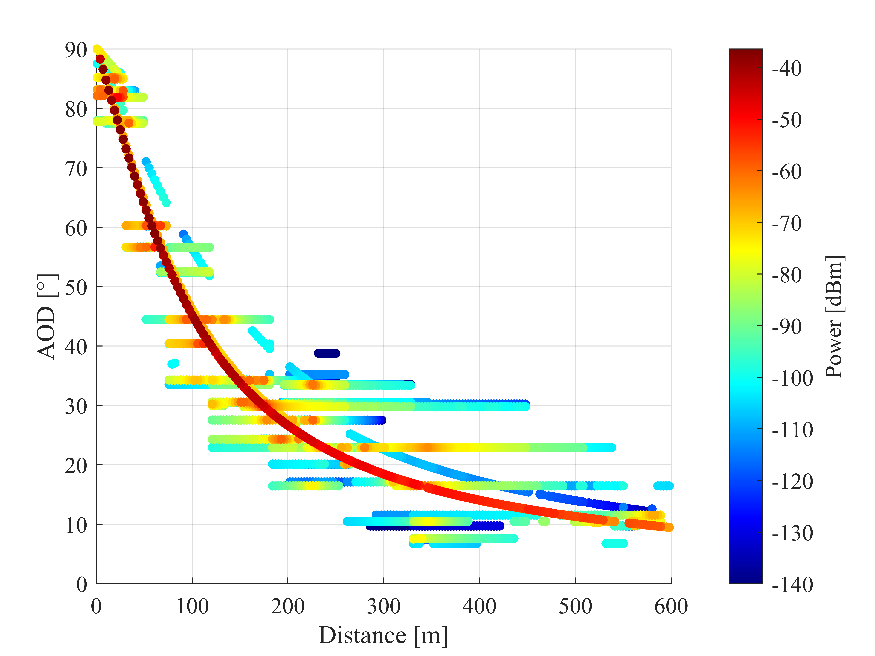}%
\label{fig_first_case}}
\hfil
\subfloat[]{\includegraphics[width=0.48\linewidth]{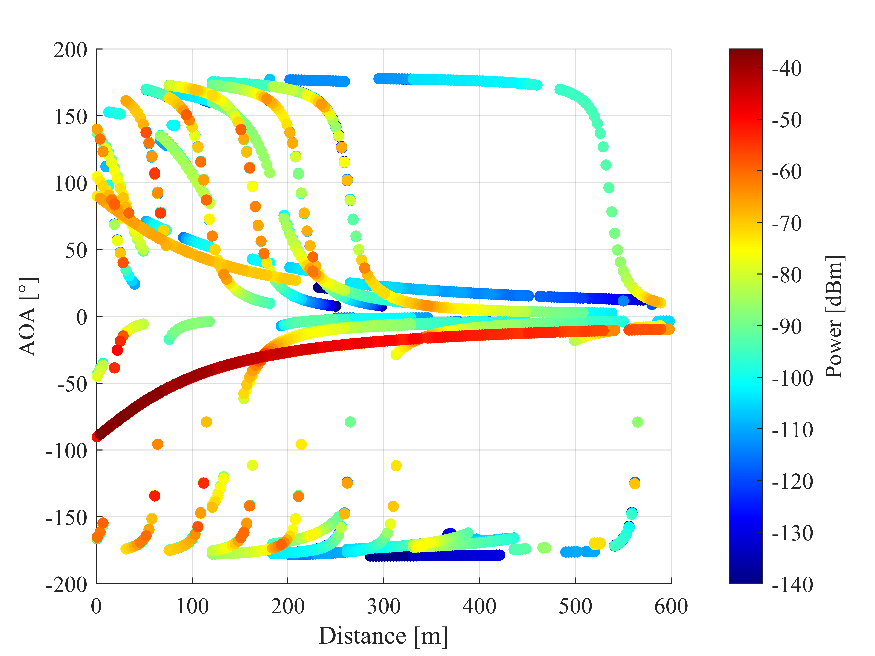}%
\label{fig_second_case}}
\caption{PAS results in the cutting scenario with narrow-beam antenna (Type C). a) AOD. b) AOA.}
\label{fig_sim}
\vspace{-0.2cm}
\end{figure*}

\begin{figure}[!t]
\centering
\vspace{-0.45cm}
\includegraphics[width=3.5in]{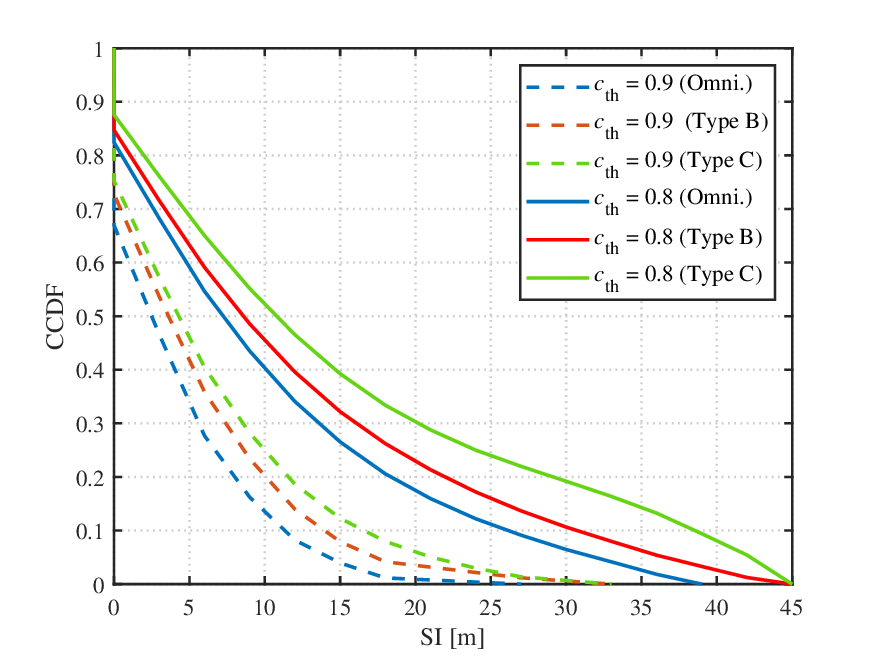}
\caption{CCDF results of SI in the station scenario.}
\label{fig9}
\vspace{-0.45cm}
\end{figure}

\vspace{-0.3cm}
\subsection{Non-Stationary Characteristics}

The local region of stationarity (LRS) method is employed to estimate the SI, which identifies the longest duration or separation where the correlation coefficient between successive PDPs remains above a specified threshold ($c_{th}$) [34]. Considering $c_{th} = 0.8$ and $c_{th} = 0.9$, complementary cumulative distribution function (CCDF) results of SI in the station scenario with three kinds of antenna configuration are presented, as shown in Fig. 9. It can be seen that as the $c_{th}$ decreases, the correlation threshold across different snapshots become more lenient, leading to an increase in SI. Table III lists the mean value results of SI in the three HSR scenarios. It can be observed that at the same beamwidth  and threshold, the station scenario exhibits the shortest SI, followed by the cutting scenario. One reason for this phenomenon is the presence of numerous obstructions (such as canopies and columns) in station environments, which results in greater non-stationarity under identical conditions. In contrast, the viaduct scenario demonstrates significantly longer SI compared to the station and cutting scenarios. This is because the viaduct scenario has fewer obstructions, with the majority being LOS components, leading to less pronounced temporal variations in the channel. On the other hand, as the beamwidth increases, the SI decreases. This occurs because wider beams cover more scatterers, resulting in stronger independence of channel parameters within each snapshot and consequently shorter SI.

\begin{table}[t]
  \centering
  \renewcommand\arraystretch{1.1}
  \small\caption{Mean value results of SI in simulated HSR scenarios}
  \label{table3}
  \begin{tabular}{c|c|cc}
    \hline\hline
    \multirow{2}{*}{\textbf{Scenario}} & \multirow{2}{*}{\textbf{Beam Type}} & \multicolumn{2}{c}{\textbf{SI [m]}} \\
    & & $c_{\text{th}} = 0.8$ & $c_{\text{th}} = 0.9$ \\
    \hline
    \multirow{3}{*}{Viaduct} & Omni. & 22.5 & 6.0 \\
    & Type B & 33.9 & 9.2 \\
    & Type C & 54.0 & 11.5 \\
    \hline
    \multirow{3}{*}{Cutting} & Omni. & 10.5 & 4.4 \\
    & Type B & 11.6 & 4.9 \\
    & Type C & 17.1 & 6.6 \\
    \hline
    \multirow{3}{*}{Station} & Omni. & 6.2 & 3.1 \\
    & Type B & 9.8 & 3.8 \\
    & Type C & 12.4 & 5.9 \\
    \hline
  \end{tabular}
\vspace{-0.45cm}
\end{table}

\vspace{-0.25cm}
\section{Performance Evaluation for 5G-R Systems in Narrow-Beam Channels}

\subsection{Performance Evaluation Based on Vienna 5G Simulator}

To investigate the performance of 5G-R systems in typical HSR scenarios with narrow-beam channels, the Vienna 5G simulator [35] is employed, which enables system-level simulations of 3GPP protocols and evaluates the performance at both physical and higher layers. Additionally, the simulator supports flexible modification of simulation parameters, allowing users to create diverse simulation scenarios and import custom channel models. This makes it highly suitable for performance evaluation for 5G-R systems in narrow-beam channels.

The system parameters are configured according to the 5G-R frequency plan of China: uplink at 1965–1975 MHz, downlink at 2155–2165 MHz, a subcarrier spacing (SCS) of 15 kHz, a slot duration of 1 ms, 14 orthogonal frequency-division multiplexing (OFDM) symbols per slot, and 10 slots per transmission block (one frame). The total number of simulated transmission blocks is 5000. The simulation assumes a high-speed train with a total length of 100 m and a width of 3.5 m, carrying 300 passengers inside the train. The passenger distribution follows a Poisson distribution with a density of $5\times10^{-5}$ at their respective locations. The simulation scenario refers to a multi-BS scenario with an inter-site distance of 2000 m, and the other scenario settings are basically the same as that of RT-based narrow-beam channel simulation. The channel parameters are configured according to the RT-based narrow-beam channel simulation results in the viaduct scenario with directional antenna (Type C). Based on the Vienna 5G simulator, the performance of the 5G-R system is evaluated in terms of BLER, throughput and spectral efficiency.

Fig. 10 shows the BLER versus distance curves under the narrow-beam channel (Type C) at train speeds of 200 km/h, 300 km/h, and 350 km/h, respectively. It can be seen that in case of narrow-beam channel, varying train velocity has minimal impact on the BLER. This is attributed to the spatial filtering effect of narrow-beam antennas, which reduces the MPCs and concentrates the AOA distribution. Consequently, speed variations within this range do not significantly alter Doppler effects. Furthermore, the beam tracking mechanism combined with high transmit gain from narrow-beam antennas ensures strong received signal power and high SINR at the RX. These two factors collectively make BLER relatively insensitive to train speed variations. It can be also found that within the 0-500 m and 1500-2000 m ranges, the BLER fluctuates markedly between 0.1 and 0.9. This occurs because these regions represent cell-edge areas between adjacent BSs where received signal power is low due to distance-dependent PL and Doppler shift reaches its maximum. In addition, the BLER decreases sharply between 500-1500 m, reaching its minimum at 700 m and maintaining stability through 700-1300 m. This optimization results from improved channel quality due to increased received power.

\begin{figure}[!t]
\centering
\vspace{-0.3cm}
\includegraphics[width=3.5in]{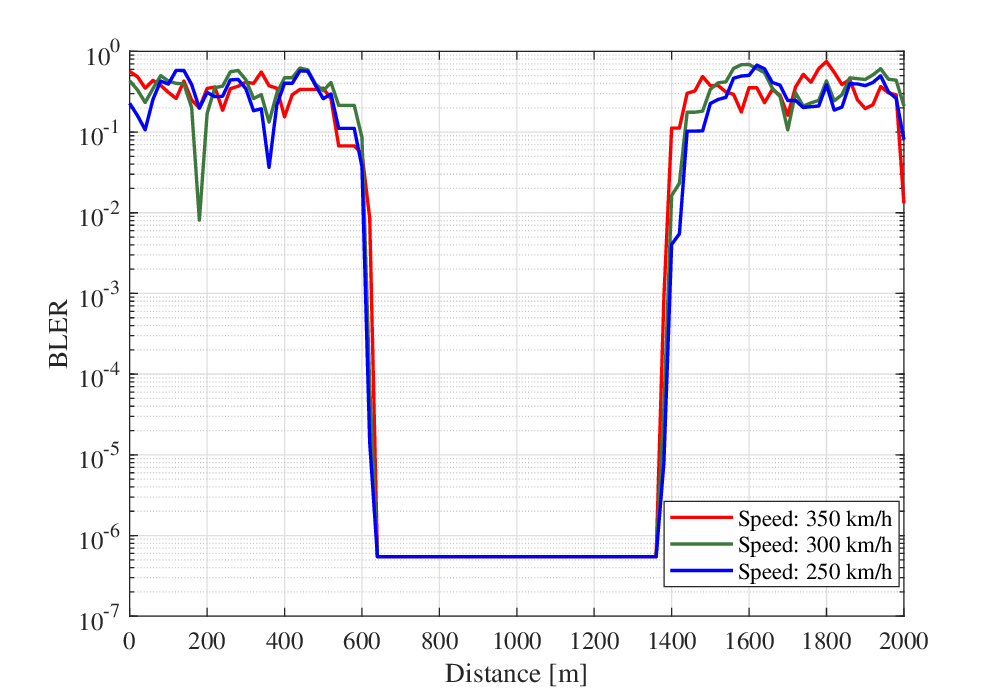}
\caption{BLER results under the narrow-beam channel (Type C).}
\label{fig10}
\vspace{-0.6cm}
\end{figure}

The variation of instantaneous throughput under different channel types and moving speeds is illustrated in Fig.11. As the train travels from 0 m to 1000 m, the throughput gradually increases. When the train approaches the BS, the throughput curve flattens out. As the train moves away from the BS, the SINR deteriorates and BLER increases, leading to a significant drop in throughput. When the omnidirectional channel is considered in simulations, the system throughput at 350 km/h exhibits the lowest average value, while the throughput at 250 km/h is the highest. Moreover, the impact of speed variation on throughput becomes less pronounced when the train is closer to the BS. This is because higher speeds introduce greater Doppler shifts, resulting in increased inter-carrier interference (ICI) and reduced SINR, thereby degrading throughput. However, near the BS, since the AoA of MPC and the direction of the train’s motion form a specific angle, e.g., 90°, the Doppler shift becomes zero, allowing the throughput to reach its peak regardless of speed. On the other hand, the throughput of narrow-beam channel shows little sensitivity to changes in train speed. This is because narrower beamwidth reduces frequency dispersion in the channel, making the narrow-beam channel less susceptible to Doppler shifts. Consequently, variations in speed have a minimal impact on throughput performance.

\begin{figure}[!t]
\vspace{0.15cm}
\centering
\includegraphics[width=3.5in]{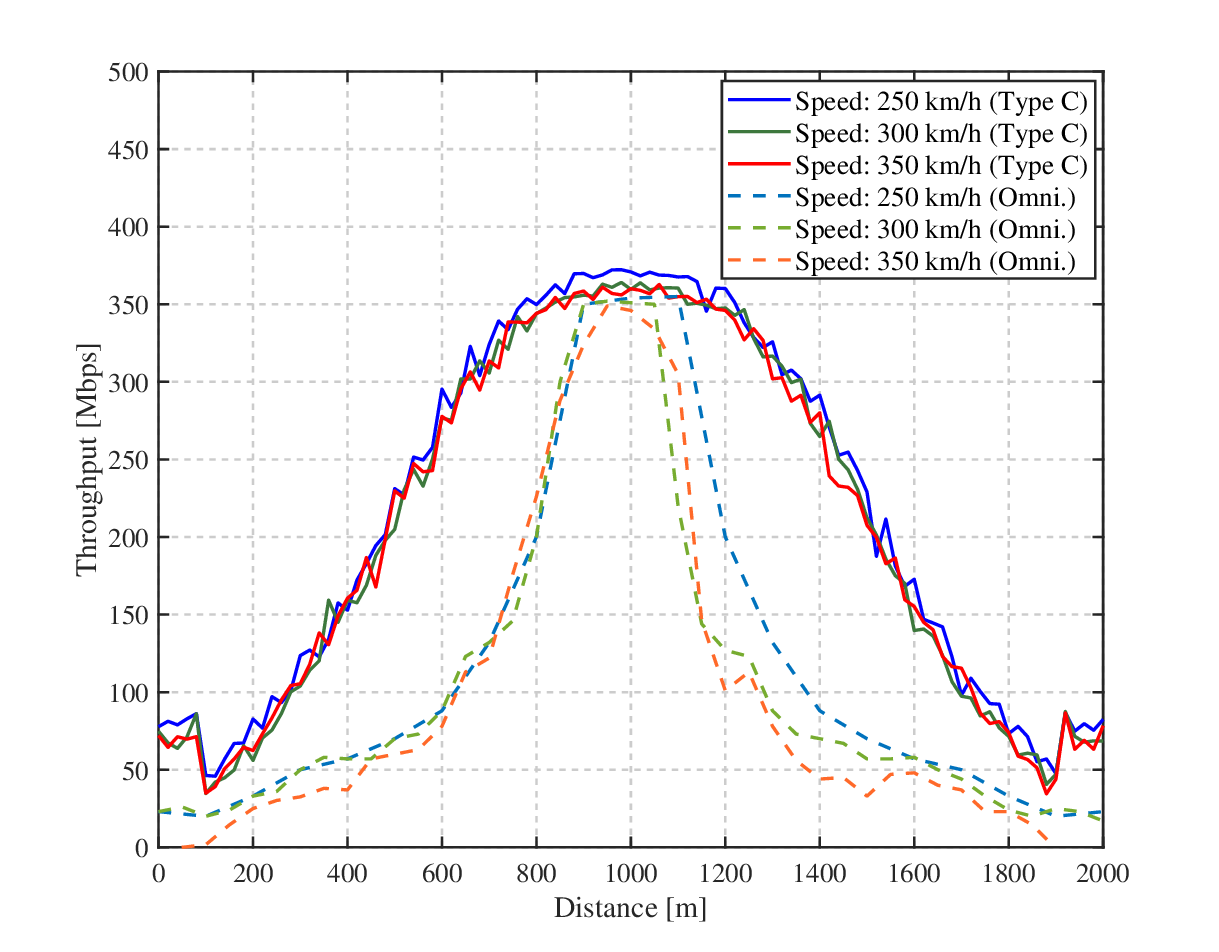}
\caption{Throughput results under different channels.}
\label{fig11}
\vspace{-0.5cm}
\end{figure}

Fig. 12 presents the inverse cumulative distribution function (ICDF) of spectral efficiency under different channels and varying mobility speeds. The results demonstrate that the average spectral efficiency of the omnidirectional channel is approximately 5 bps/Hz, whereas that of the narrow-beam channel reaches around 23 bps/Hz. This performance enhancement can be attributed to the factor that the narrow-beam provide substantial radiation gain at the TX and effectively increases the SINR at the RX, allowing more effective data blocks to be transmitted per unit time within the same bandwidth. Furthermore, it reveals that the spectral efficiency decreases as terminal velocity increases. This degradation effect is particularly pronounced in the omnidirectional channel.

\begin{figure}[!t]
\centering
\vspace{-0.3cm}
\includegraphics[width=3.5in]{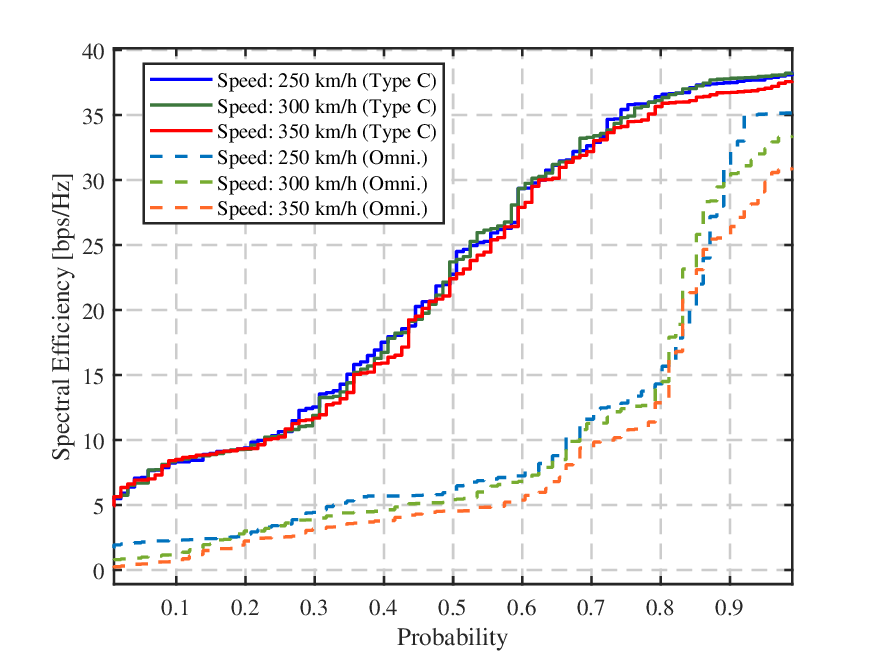}
\caption{ICDF results of spectral efficiency under different channels.}
\label{fig12}
\vspace{-0.45cm}
\end{figure}

\vspace{-0.45cm}
\subsection{Performance Evaluation Based on HIL Simulation Platform}

To further evaluate the performance of 5G-R systems, we establish a 5G-R HIL simulation platform, as shown in Fig. 13. The HIL simulation platform includes Propsim F32 channel emulator, 5G-R BS, 5G-R core network, radio block center of train control system (RBC), user equipment (UE), and a laptop equipped with signal processing software. The Propsim F32 enables full-stack, end-to-end performance testing of wireless devices, radio frequency (RF) systems, and network infrastructure equipment in a laboratory environment. In addition to the channel emulator, hardware used in the HIL simulation platform is the same as that of the 5G-R network equipment which will be deployed on HSR in China. The UE employs a communication module developed in the laboratory, which integrates hardware components such as 5G baseband chip, RF module, memory, power management into a single package, while providing standardized software and hardware interfaces for external connectivity. The 5G communication module, combined with a cellular drive test system (CDS) software, enables data acquisition, storage, analysis, positioning and reporting for the 5G-R network.

\begin{figure*}[!t]
\centering
\vspace{-0.2cm}
\includegraphics[width=5in]{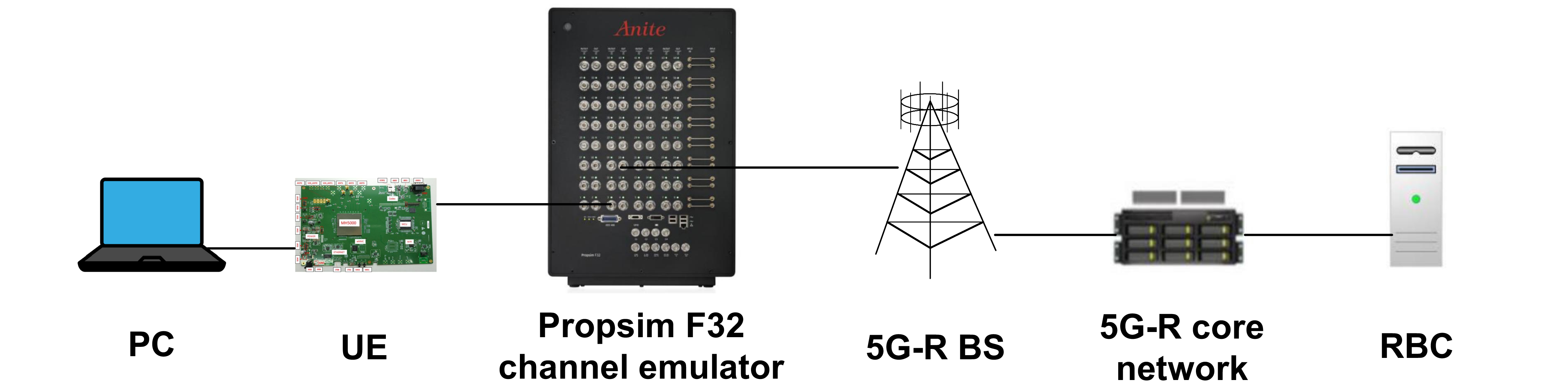}
\caption{Schematic diagram of the 5G-R HIL simulation platform.}
\label{fig13}
\vspace{-0.2cm}
\end{figure*}

Using the 5G-R HIL simulation platform, we conduct the performance evaluation for 5G-R systems, as shown in Fig. 14. The system parameter settings are basically the same as that of the Vienna 5G simulator-based performance evaluation. The simulation scenario is configured with an inter-BS distance of 700 m, with the train entering the coverage area of the next BS approximately every 10 seconds. It is noted that the train operates at a constant speed of 250 km/h throughout the entire simulation period. In the channel emulator, the large-scale and small-scale channel parameters are configured according to the RT-based narrow-beam channel simulation results in the viaduct scenario with different antennas. Based on the 5G-R HIL simulation platform, the performance of the 5G-R system is evaluated by SS RSRP, SINR and RSRQ. SS RSRP is a key indicator for evaluating signal coverage in mobile communication systems. SS RSRQ, derived from SS RSRP, serves as an important reference for assessing received signal quality in 5G communication systems and determines cell reselection and cell handover.

\begin{figure}[!t]
\vspace{-0.2cm}
\centering
\includegraphics[width=3in]{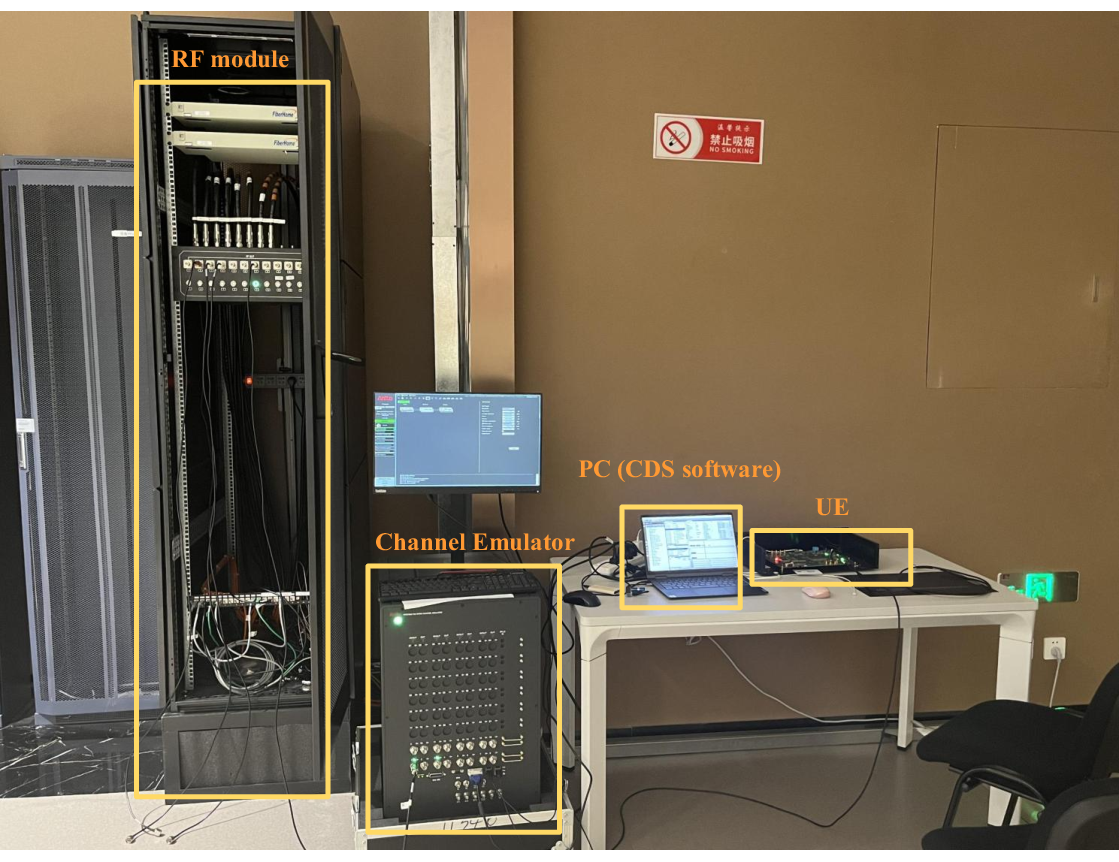}
\caption{Test environment of the 5G-R HIL simulation platform.}
\label{fig14}
\vspace{-0.4cm}
\end{figure}

The simulation results of SS RSRP temporal variation within 0-40 s are shown in Fig.15 The plot demonstrates periodic fluctuations in RSRP, which decreases as the train moves away from the BS, reaching its minimum after 5 seconds. Comparative analysis for the three channels reveals that the narrow-beam channel achieves the highest average power, while the omnidirectional channel yields the lowest values. The narrow-beam solution demonstrates superior performance compared to the wide-beam and omnidirectional solutions, with trains spending most operational time in high-coverage-strength regions. However, at cell edges, the performance of the narrow-beam solution degrades below that of the wide beam solution, as the additional PL for narrow-beam channel in far-field scenarios outweighs the antenna gain benefits.

\begin{figure}[!t]
\vspace{-0.6cm}
\centering
\includegraphics[width=3.5in]{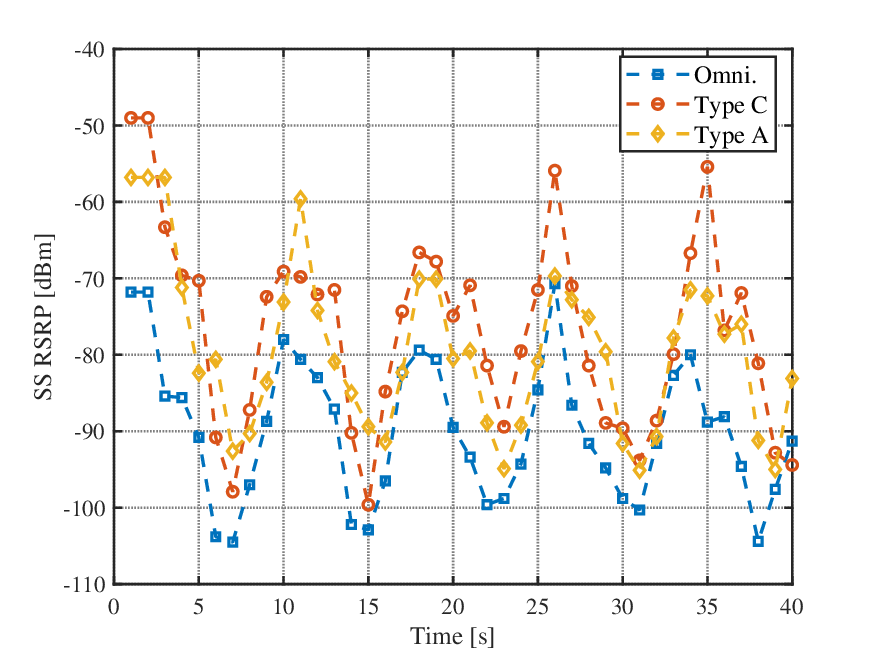}
\caption{SS RSRP results under different channels.}
\label{fig15}
\vspace{-0.15cm}
\end{figure}

\begin{figure}[!t]
\centering
\vspace{-0.3cm}
\includegraphics[width=3.5in]{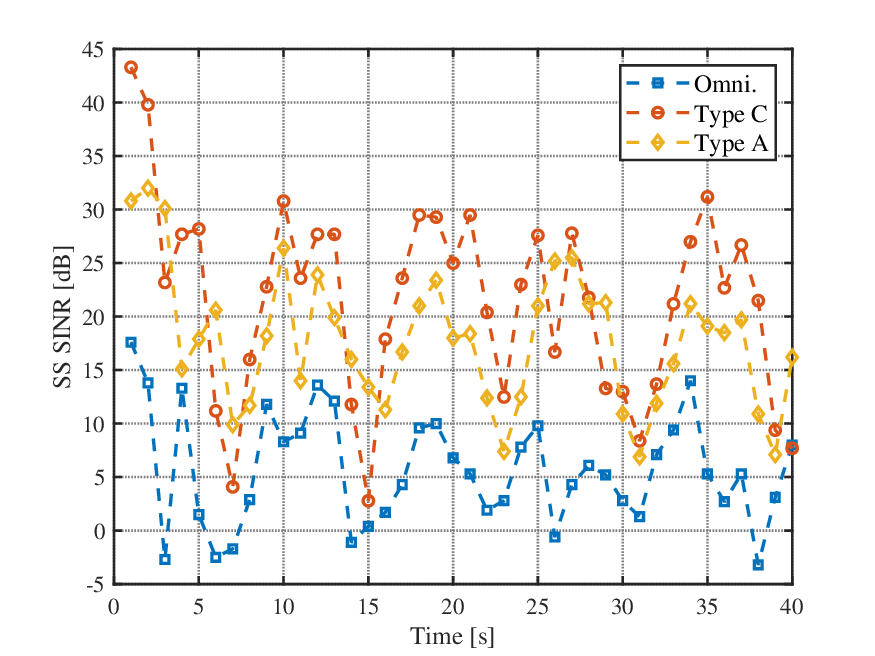}
\caption{SS SINR results under different channels.}
\label{fig16}
\vspace{-0.6cm}
\end{figure}

Fig. 16 presents the variation curve of SS SINR versus travel time. As the beamwidth decreases, the variation trend of SINR becomes similar to that of RSRP. However, at far coverage points, the SINR performance of the narrow-beam channel approaches or even underperforms the wide-beam configuration. This phenomenon occurs because the additional PL introduced by the narrow beam at cell edges outweighs the benefits of antenna gain, resulting in inferior SINR performance compared to the wide-beam antenna.

Fig. 17 illustrates the RSRQ variation over time. The results demonstrate that the narrow-beam and wide-beam channels maintain excellent RSRQ values ranging from $-$10.8 dB to $-$11 dB, indicating superior received signal quality. In comparison, the omnidirectional channel shows relatively poorer RSRQ performance with noticeable location-dependent variations. As the train moves away from the BS, the RSRQ decreases rapidly but remains better than $-$12 dB even at the worst-case position, suggesting maintained good communication quality.

\begin{figure}[!t]
\centering
\vspace{-0.45cm}
\includegraphics[width=3.5in]{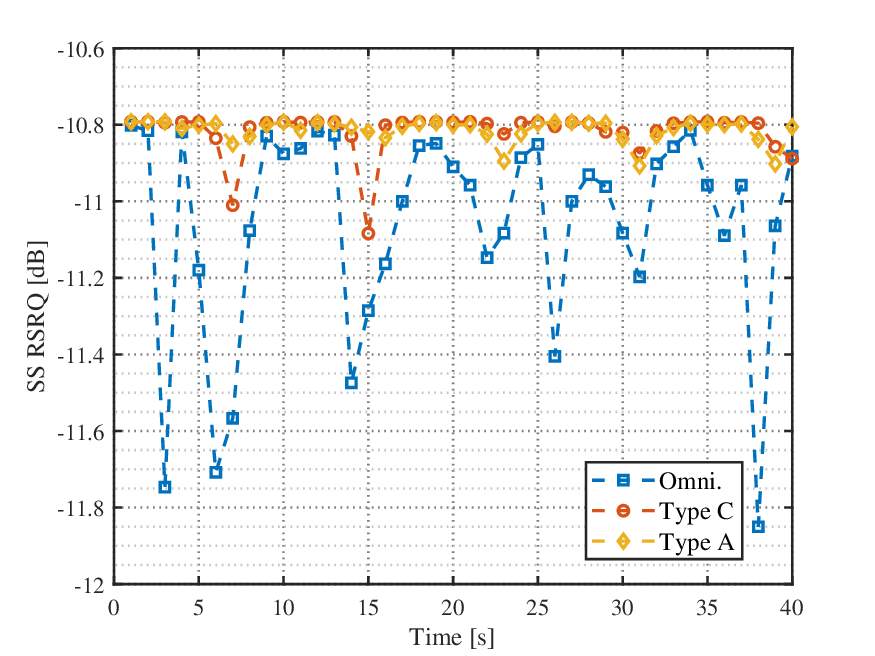}
\caption{SS RSRQ results under different channels.}
\label{fig17}
\vspace{-0.45cm}
\end{figure}

\vspace{-0.3cm}
\section{Conclusion}

This paper has investigated the narrow-beam channel characteristics and performance of 5G-R systems through RT simulations. By establishing the three typical HSR scenarios such as viaduct, cutting, and station and performing the RT-based dynamic narrow-beam channel simulation, we have obtained comprehensive statistical insights into both large-scale and small-scale fading properties, as well as channel non-stationarity. Key parameters including PL, SF, fading severity, time-frequency-space dispersion, and SI have been quantitatively analyzed, with particular attention given to the impact of beamwidth on channel behavior. Furthermore, the performance of 5G-R systems in the narrow-beam channel has been evaluated based on the Vienna 5G simulator, focusing on BLER, throughput and spectral efficiency. Additionally, the 5G-R HIL simulation platform has been developed and utilized to assess the performance of 5G-R systems in terms of SS RSRP, SINR and RSRQ. These results will provide helpful information for the design and optimization of 5G-R systems in HSR scenarios.

\end{document}